# Field Measurement for Superconducting Magnets of ADS Injector I

YANG Xiang-chen (杨向臣), PENG Quan-ling (彭全岭)
Institute of High Energy Physics, Chinese Academy of Sciences, Beijing 100049, China

**Abstract**

The superconducting solenoid magnet prototype for ADS injection-I had been fabricated in Beijing Qihuan Mechanical and Electric Engineer Company and tested in Haerbin Institute of Technology (HIT) in Nov, 2012. Batch magnet production was processed after some major revision from the magnet prototype, they include: removing off the perm-alloy shield, extending the iron yoke, using thin superconducting cable, etc. The first one of the batch magnets was tested in the vertical Dewar in HIT in Sept. 2013. Field measurement was carried out at the same time by the measurement platform that seated on the top of the vertical Dewar. This paper will present the field measurement system design, measurement results and discussion on the residual field from the persistent current effect.

## 1. Overview of ADS superconducting magnet

The Accelerator Driven Sub-critical System (ADS) is based on the superconducting accelerating structure, the purpose of which is to realize the safe disposal of nuclear waste [1]. The lattice structure for the injection-I was designed by IHEP and described somewhere [2]. Physical design of the superconducting magnet prototype for ADS was designed by IHEP and tested in HIT [3]. Each superconducting magnet, aimed for beam focusing and beam orbit correction, contains a solenoid magnet, a horizontal dipole corrector (HDC) and a vertical dipole corrector (VDC). The designed integral field for the solenoid is 0.4T-m, and the maximum integral field for HDC and VDC is 1600G-cm. Figure 1 shows the cross section of the magnet prototype, the solenoid field is realized by the main solenoid coil, the two bucking solenoid coil, and the iron return yoke, an extra perm-alloy tube covering the iron yoke is used to reduce the leakage field further. The final design value for the magnet prototype, the leakage fields at 270mm away from the solenoid center are 0.045G with perm-alloy shield and 0.3G without perm-alloy. The two correctors are wound inside the main solenoid coil, they are selected as saddle shaped coils in order to save space and hence to reduce the total store energy of the magnet. The support tube, where the superconducting coils wound on, also serves as the beam vacuum chamber and as the inner helium vessel. .

In order to know the magnetic field performance at several operation currents, a vertical test system was fabricated in 2012, it had been used to test the magnet prototype and the first one of the batch magnet in Nov. 2012 and Sept. 2013 respectively [3].

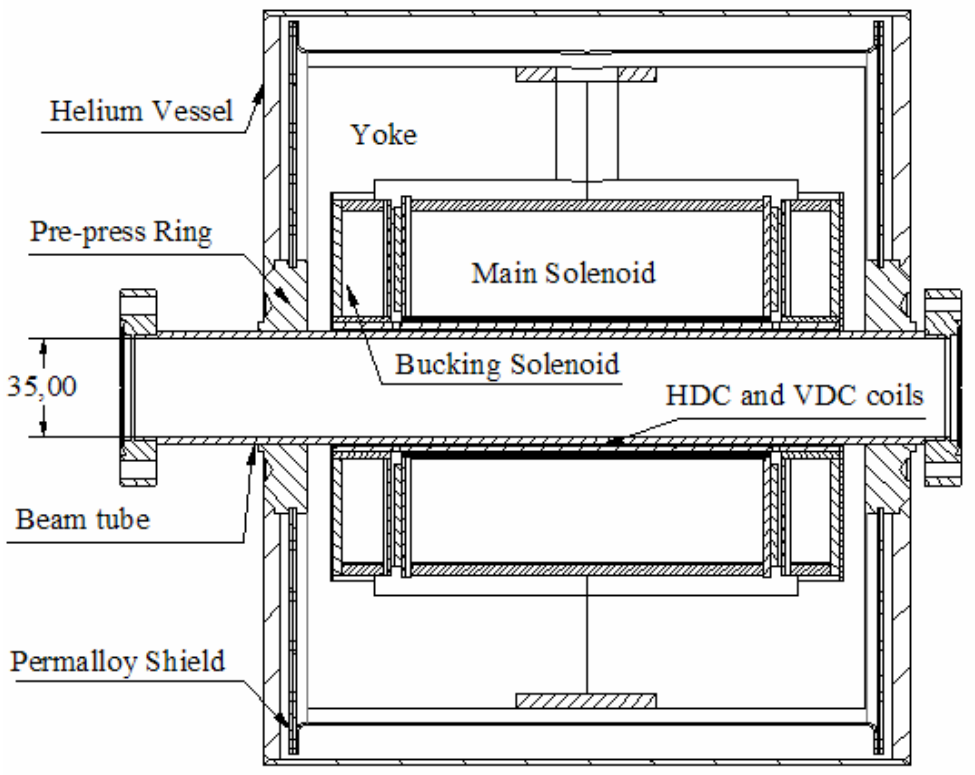

Fig. 1. Schematic overview of the superconducting solenoid magnet prototype

## 2. Vertical test system

### 2.1 Overview of the vertical vest dewar and the measurement platform

Figure 2 shows the assembly structure of the top flange components, on the top of the top flange the moveable measurement platform is seated on. The bare superconducting magnet, where the outer helium vessel not welded on, will be fixed on the hoisting flange vertically. The isolation tube, which is really a two concentric 316L stainless tubes to form a vacuum space in between, can realize the thermal isolation from liquid helium state to room temperature state. The superconducting magnet should be installed on the hoisting flange vertically and keep support tube with the isolation tube coaxially. Figure 3 shows the schematic layout of the field measurement platform. A three axis Hall probe, which is embedded at the end of the measuring tube, can move vertically inside the isolation tube with a maximum distance of 460mm. The movement module, which as a whole is driven by a servo motor, could pull the measuring tube inside the isolation tube up and down with the setting steps. To keep the isolation tube and the measuring tube coaxially, the measurement platform can be adjusted in horizontal by manual. Another function of the measurement platform is that it can realize the rotating field measurement when measurement tube is replaced by a rotating coil.

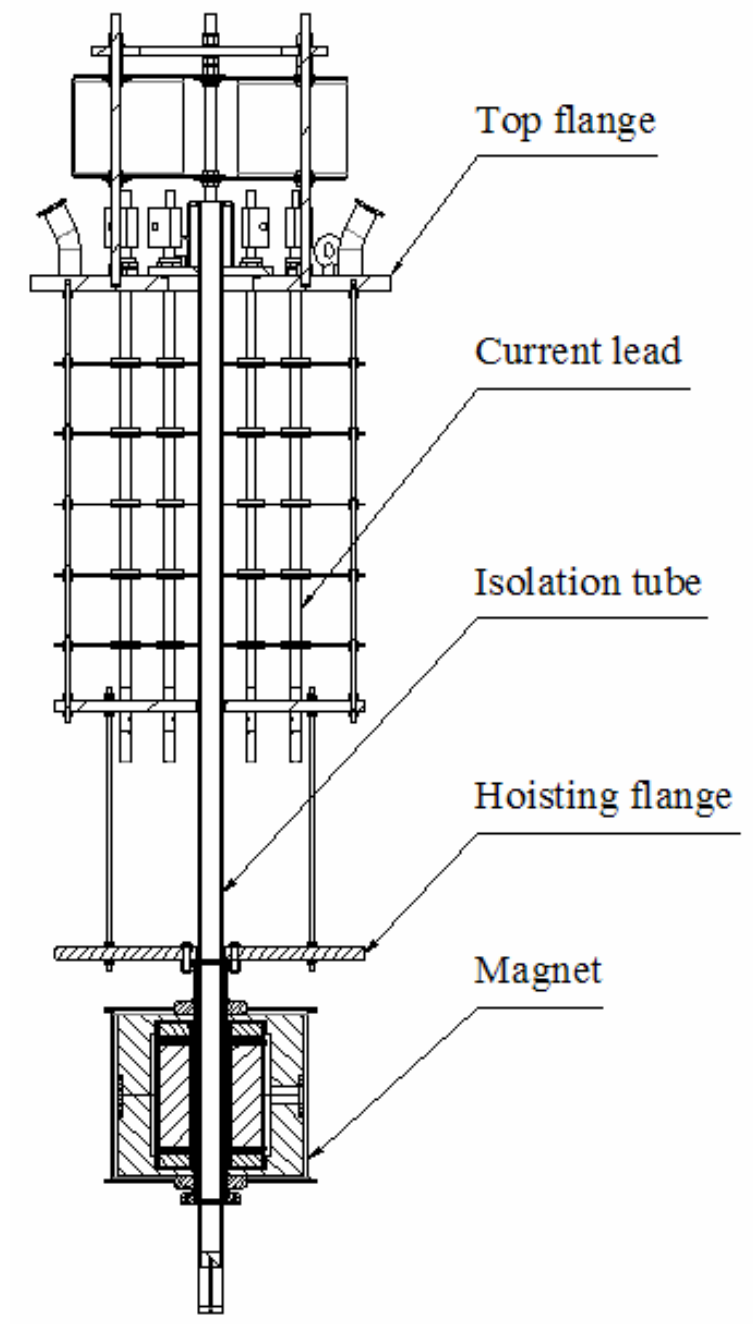

Fig. 2. The assembly structure of the top flange components

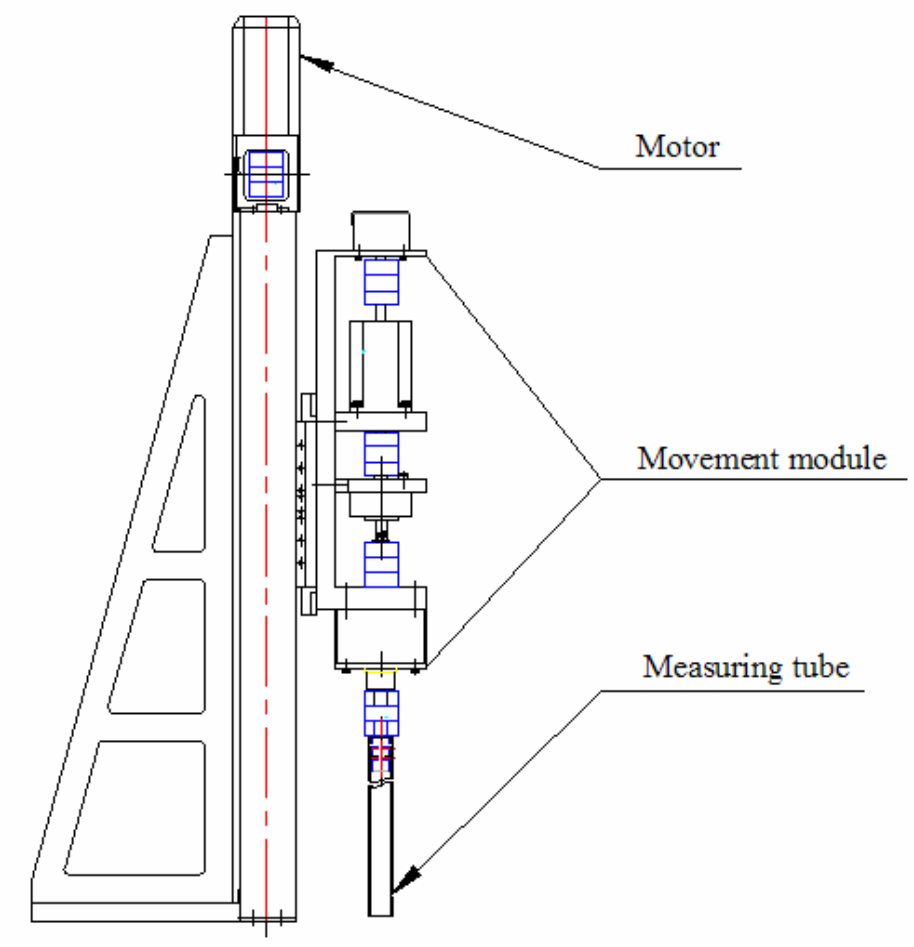

Fig. 3. Schematic layout of the field measurement platform

### 2.2. Hardware of the measurement system

The measurement system is realized in a half-closed loop control. A grating ruler, which is glued on the side of the measurement paltform, can give the moving distance in a precision of 0.5 microns. The measurement platform was controlled by EMAC controller, it communicates with computer via Ethernet cable. The movement orders, which are sent from computer to the EMAC controller, are then amplified by the servo driver and sent to the servo motor to pull the three axis Hall probe to move vertically. A domestic made CH-3600 gauss meter, with the accuracy of 0.15% and measurement range of 10T, has been calibrated with temperature compensation. The importance of which is that the isolation tube is far below the room temperature since the cold radiation from the liquid helium in the vertical Dewar. Nitrogen gas is blown from the top of the isolation tube to get rid of ice forming inside the tube.

### 2.3. Software design of the measurement system

The measurement software was written in VC++ 6.0, which includes: system initialization, servo motor control module, data acquisition and data storage.

The system initialization is to realize the connection between computer and EMAC controller, the connection between computer and the gauss meter. The computer reads the measurement field data from the gauss meter through RS232 serial port. The total moving distance, moving distance for each step, moving velocity and acceleration of the servo motor are set all together. The minimum time delay requirement for the Gauss Meter and a stable reading for the Hall probe have driven us to set the time stop in each measurement step as 3s.

Data acquisition is carried out after system initialization. The measurement process starts from the original point which is set by the grating ruler, where the series measurement positions are made by the servo motor and read out by the grating ruler. As the Hall probe moves to the setting position, the delay program is executed, magnetic field data from the three data channels are obtained by the computer after the data reading instructions sent to the Gauss Meter. The data judgment in the software helps to ensure data collection is completed and data format are correct before moving to the next step. All the measurement data and the real positions of the Hall probe will be displayed and stored in a file in real time.

When a test line is finished, the measurement module then go back to the original position automatically by click the back to the zero button and go on to the next measurement sequence.

## 3. Field measurement results

Magnetic field distribution of the magnet is crucial to examine the magnet design and for beam commissioning. Field measurements for the first one of the batch magnets were made in HIT in Sept. 2013. Figure 4 showed the vertical test stand, where the field measurement platform is seated on top of the vertical Dewar.

In order to eliminate the influence from the cold temperature, Hall probe together with the measuring tube was put into the isolation tube for some time before the magnet excitation. After the temperature of the Hall probe was in consistent with the test environment, the measuring tube was pulled out from the vertical Dewar. The Hall probe was then covered by a perm-alloy tube and reset to zero for each channel by manual.

### 3.1. Solenoid field performance

Quench performance test showed that the operation current of solenoid magnet can finally reach above 300A after three times of natural quench (260A, 268A, 308A) during the current ramping. The combined field excitation was successfully withstood the solenoid operated at 230A, the two corrector operated at 20A. Field measurement test were going at 50A, 100A, 150A, 182A, 210A for the solenoid in the moving distance of 460mm with 230 steps. Figure 5 showed the central field $B_0$ and the integral field for the solenoid at the measurement currents. It was found that the locations of the maximum field at different currents were not in consistent, the maximum difference was about 5mm at the low current compare with that of at the high current. It was also exhibited that the central fields nearly keep linear relationship with the currents, no saturation was found in the iron yoke for the whole current excitation process.

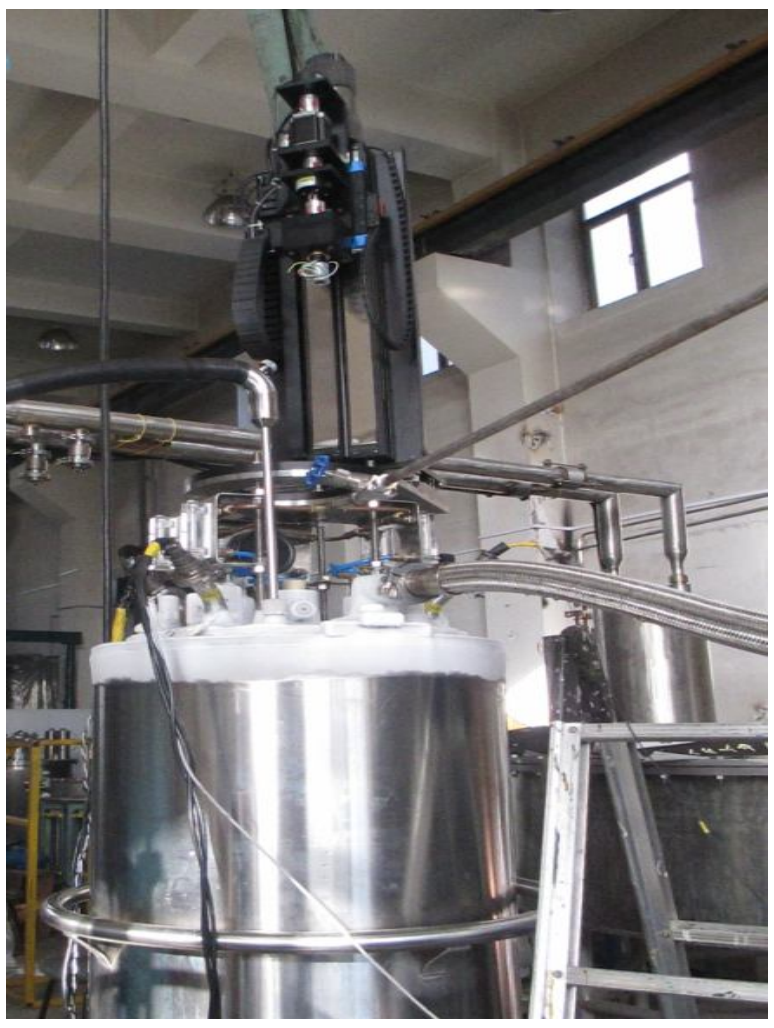

Fig. 4. Vertical test for the superconducting magnet

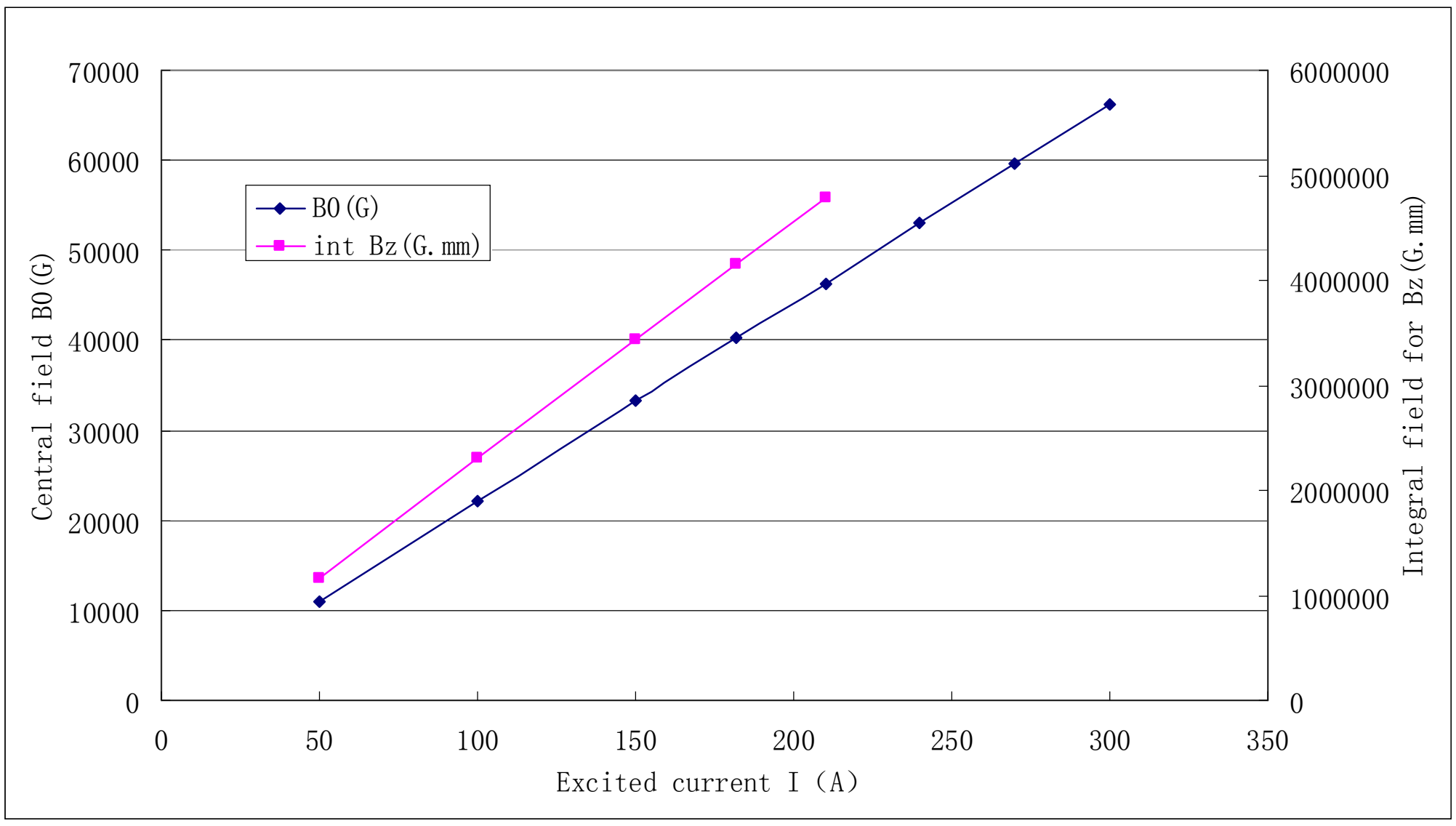

Fig. 5. Central fields and integral fields for the solenoid at several measurement currents

The designed operation current for the solenoid is 182A. In order to reduce the measurement error, the magnetic fields were measured for two times. Figure 6 shows the axial field profile at 182A. The difference between the two measurement results is less than 0.1%. The integral field is 0.416T-m. The leakage field is less than 1G at the distance of 270 mm away from the solenoid center, which meets the requirements from the upstream and downstream spoke cavities.

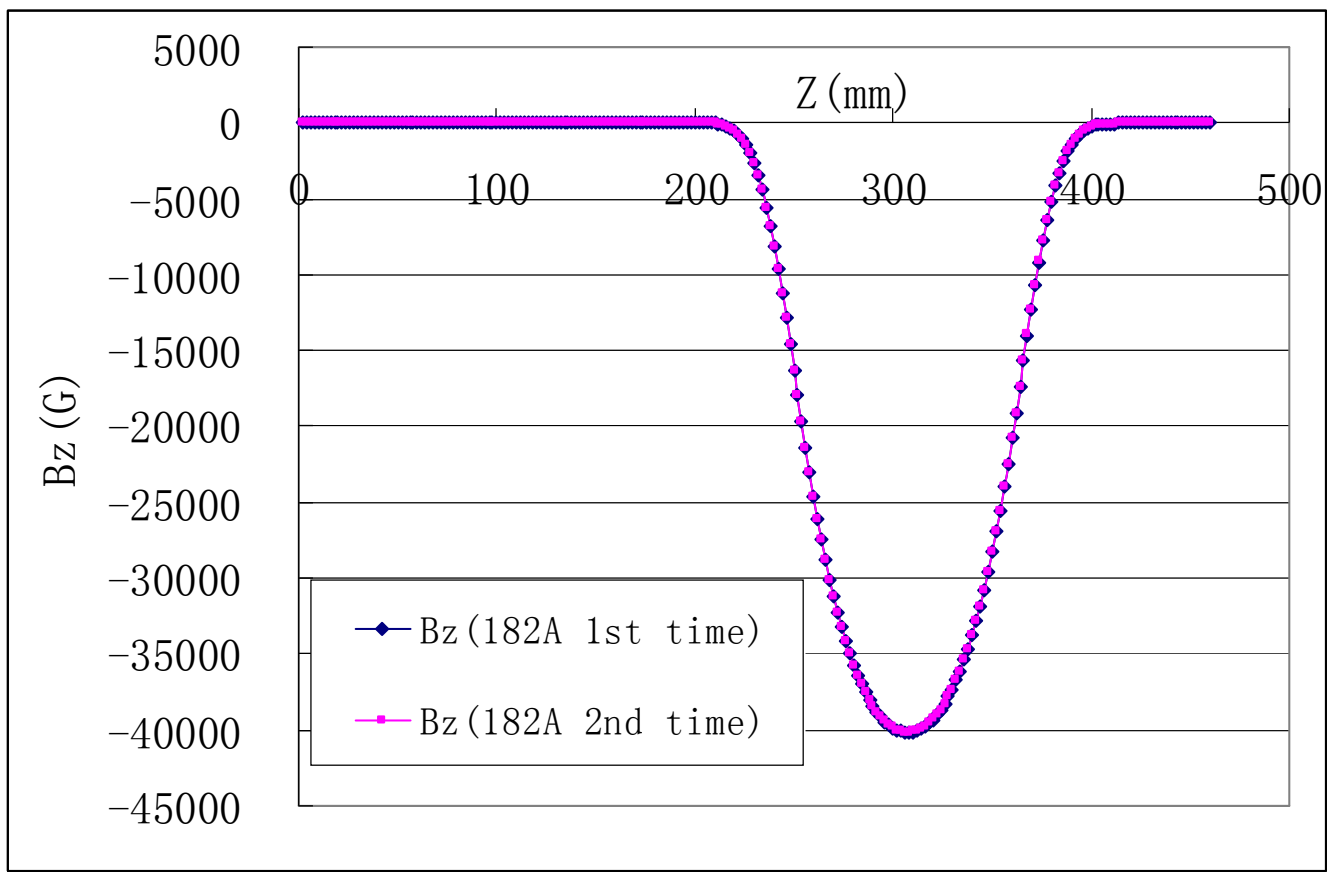

Fig. 6. Axial field (Bz) distribution for the solenoid at 182A.

**3.2. Magnetic performance of the correctors**

Figure 7 shows the field profile of the HDC at 10A, in its maximum operation current. The measured integral field is 2355.7 G-cm which is far above the designed 1600G-cm. The field profile for HDC has a longer flat shape compare with the solenoid field as shown in Figure 6 lies that saddle shaped coils for HDC and VDC correctors have covered the overall length of the main solenoid coil and the two bucking solenoid coils.

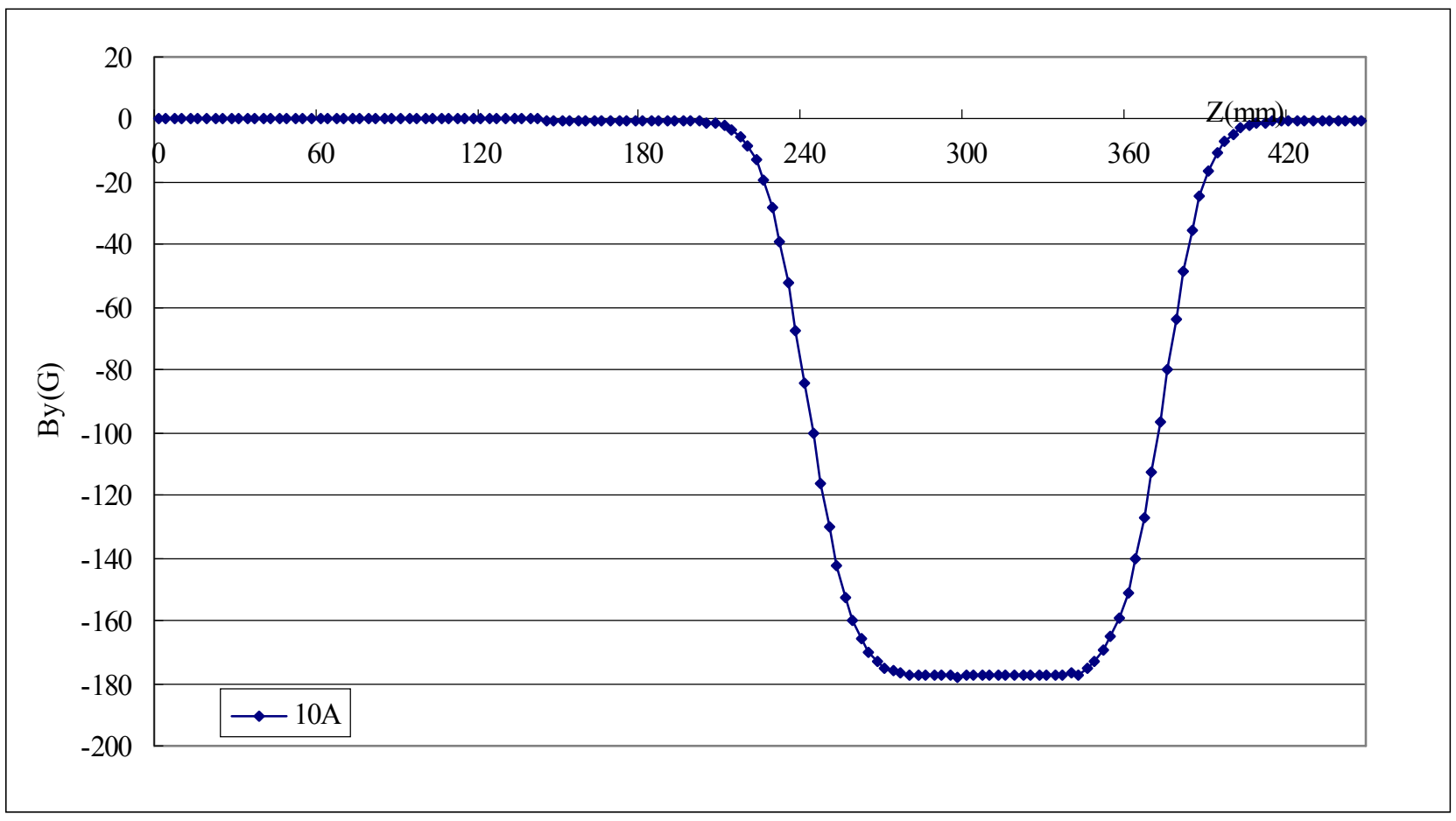

Fig. 7. Field profile of the HDC at 10A,

### 3.3. Discussion on the Residual Field from the Solenoid

Residual field of the solenoid field was measured when the solenoid current ramp down to 0A, Figure 8 shows the Residual field profile along the beam line. The maximum axial field at the main solenoid is -69G, where the peak field at the bucking solenoid is 60 G. An important issue is that the field integral for the residual field near to zero, similar as that of the HINS magnet [4]. That is to say, the magnet will affect the proton beams very limit when one of the solenoids is quench.

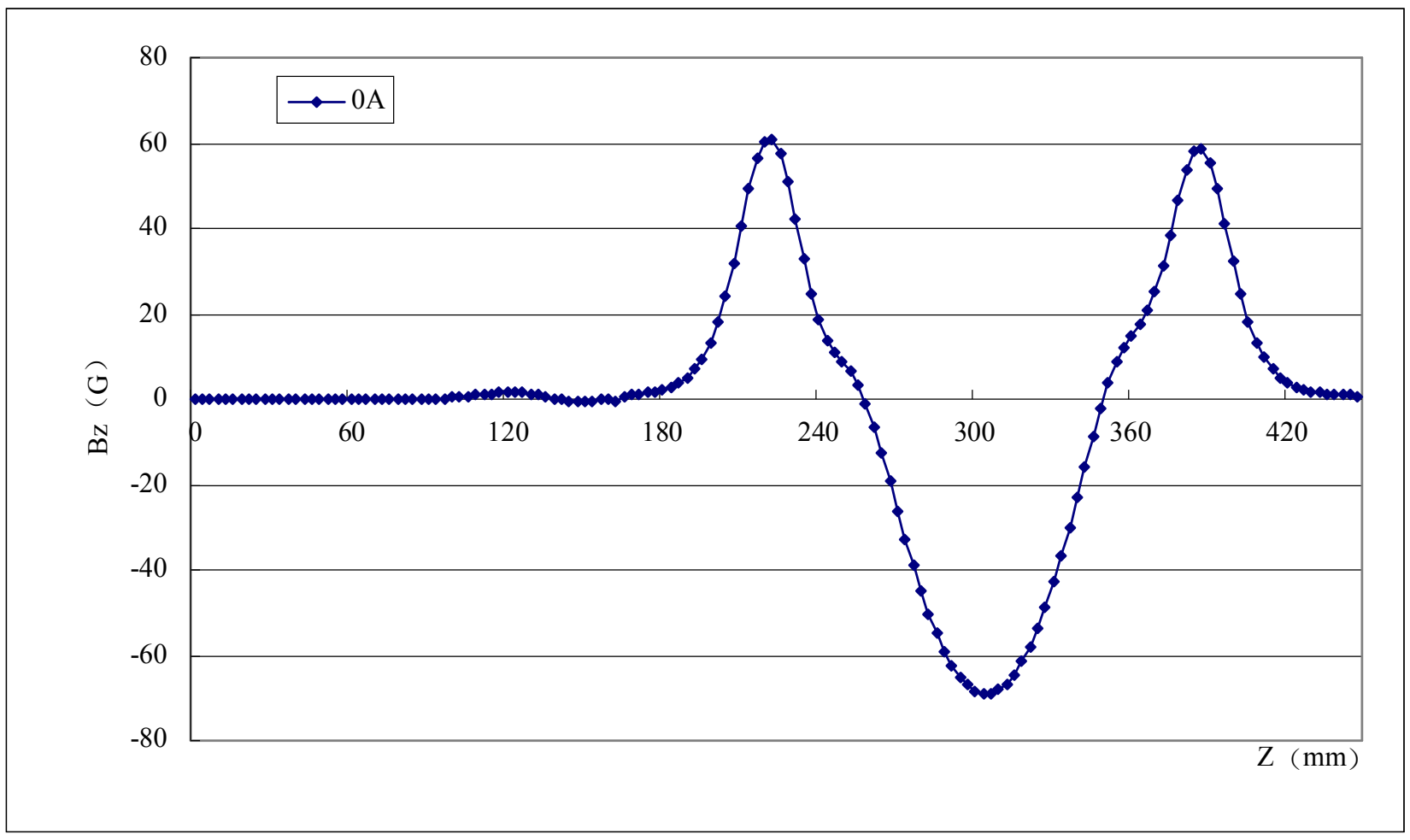

Fig. 8. Residual Field distribution of the solenoid

The superconducting cable used for the solenoids is from German EAS company, it contains 36 filaments with each 84 microns in diameter in a cross section of 1.32×0.85mm$^2$, the Cu:SC ratio is 4:1. The persistent current will be formed in each filament during current ramping process, and last several hours even if the magnet power supply is switched off [5]. Each filament will be fully magnetized by Meissner effect when the applied outer field higher than 0.66T by calculation in the coil region, which is easily met during the current ramping. Each filament will carry a current of 5.5A in its two circle halves with the opposite direction, and form a small magnetization of 0.1T. Since superconducting cable consists of 36 filaments, the field contributions from the opposite currents in the two adjacent filaments will be cancelled in the radial direction, but in axial direction their contributions by the two adjacent filaments will be added. More generally, they can be extended to more layers. For simplify, the residual field from the solenoid coil can be regarded as two thin layers carried with opposite current direction, and can be calculated in the current sheet model, similar as the field from a permanent magnet [6]. Figure 9 shows the residual field calculation model for half length of the solenoid by using OPERA-2D [7]. Figure 10 shows the calculation results, the integral field for on axis Bz nearly to zero.

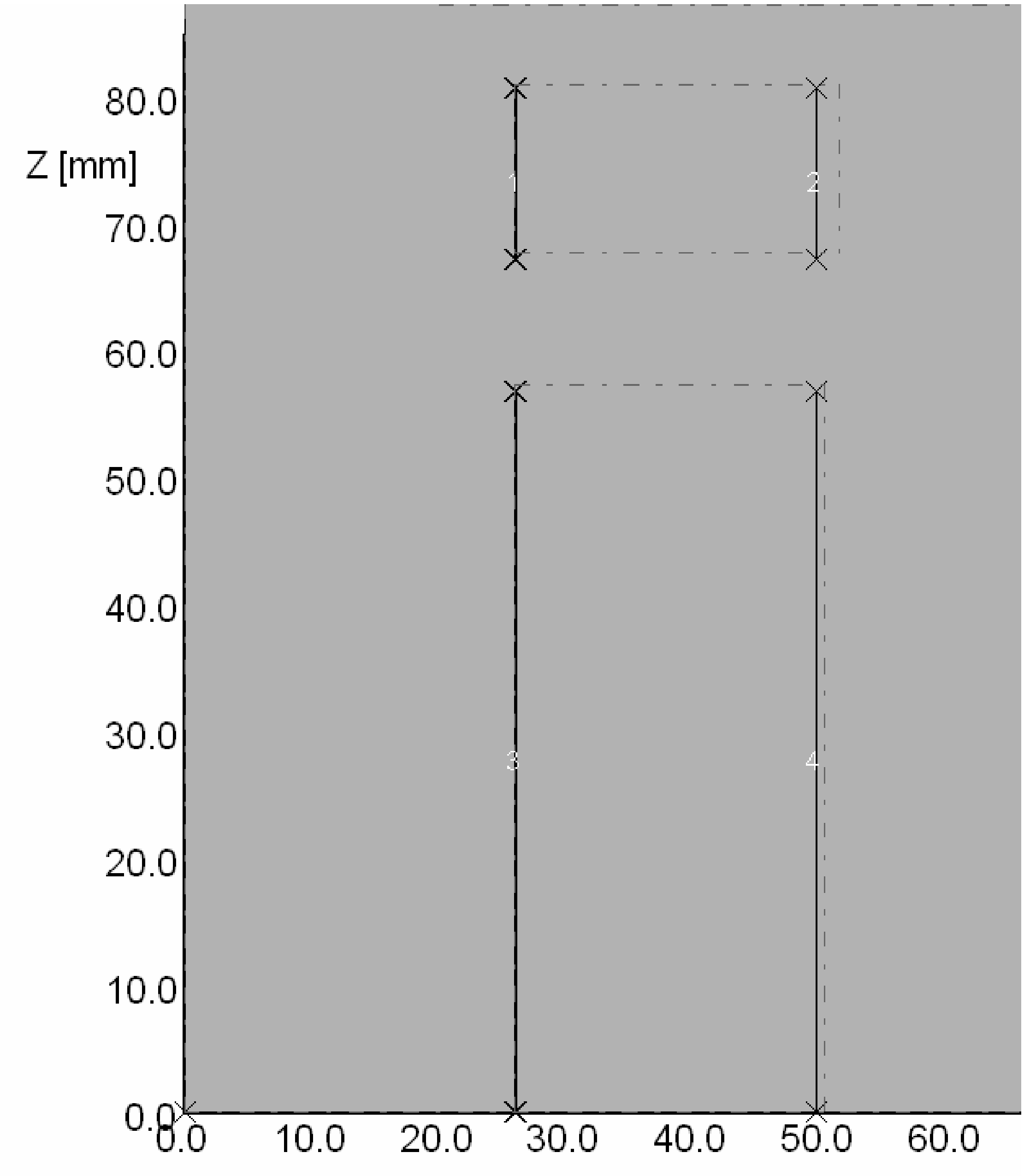

Fig. 9. OPERA-2D model for the residual field calculation. Region No. 1 to 4 are the current sheets from bucking solenoid and main solenoid respectively.

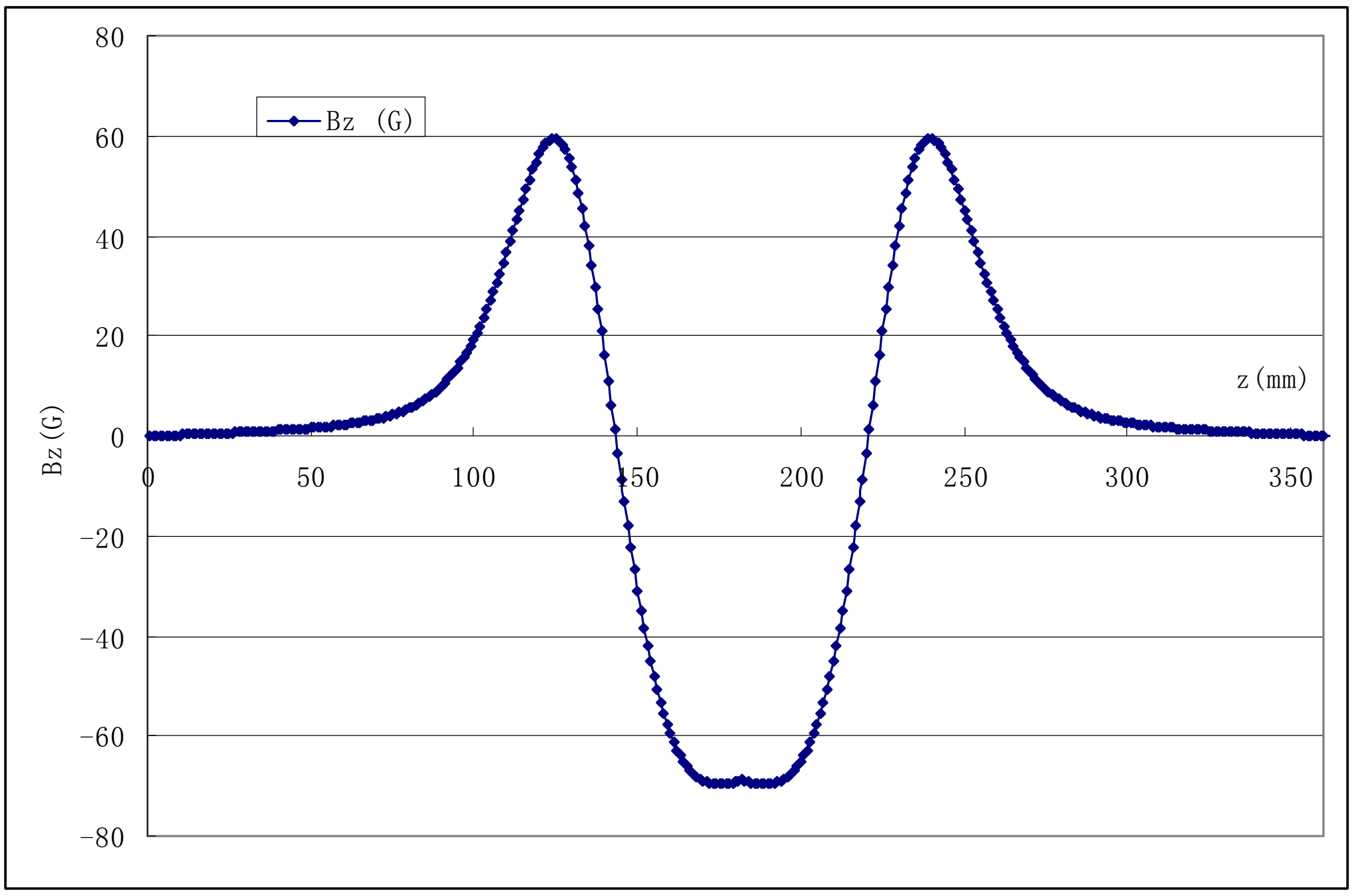

Fig. 10. Residual field calculation result for the solenoid.

## 4. Conclusion

The superconducting magnet prototype and one of the batch magnets for ADS have been tested by using the dedicated measurement platform. The measurement results agree with the design requirements in its integral field and leakage field for the solenoid magnet. Vertical test has accumulated valuable experience in quench detection and field measurements, and will give valuable directions for the subsequent horizontal test. Beyond that, study on the persistent current effect let us know the residual field analysis have no impact on the passing proton beams when one of the superconducting magnet quenched.

## References

[1] Zeller A. F., et al, IEEE Transaction on Applied Superconductivity, 2002, 12: 329.

[2] Yan Fang, Li Zhihui, Meng Cai, et al, Chinese Physics C, 2014, to be published.

[3] Peng Quan-ling, Wang Bing, Chen Yuan et al, Chinese Physics C, 2014, to be published.

[4] M. A. Tartaglia, E. Burkhardt, T. Leach, et al, IEEE Transaction on Applied Superconductivity, 2010, 20:312.

[5] S. Turner, CERN Accelerator School Superconductivity in Particle accelerators, 1989, CERN 89-04.

[6] Q. L. Peng , S. M. McMurry, J. M. D. Coey, Journal of Magnetism and Magnetic Material, 2004, 268:165.

[7] Opera Manager User Guide, Version 14R1, Vector Fields Software, July 2011